\documentclass[aps,pra,twocolumn,preprintnumbers,amssymb,amsfonts,amsmath,superscriptaddress,showpacs,hyperref,nofootinbib]{revtex4-1}

\pdfoutput=1
\usepackage{amssymb}
\usepackage{bm}
\usepackage{cancel}
\usepackage{color,graphicx}
\usepackage{amsmath}
\usepackage{relsize}
\usepackage{amsbsy}
\usepackage{amsthm}
\usepackage{bbm}
\usepackage{epsfig}
\usepackage{float}
\usepackage{graphicx}
\usepackage{dcolumn}
\usepackage{bbm}
\usepackage{color,epstopdf}
\usepackage{amscd}
\usepackage{amsfonts}  
\usepackage[]{amsmath}
\usepackage{amssymb}    
\usepackage{mathrsfs}
\usepackage{verbatim}
\usepackage[]{cases}
\usepackage{amsmath}
\usepackage{wasysym}
\usepackage[utf8]{inputenc}
\usepackage[T1]{fontenc}
\usepackage{mathtools}
\usepackage{xcolor}
\usepackage{tabularx}

\newcommand{\hide}[1]{\relax}

\newcommand{\nocontentsline}[3]{}
\newcommand{\tocless}[2]{\bgroup\let\addcontentsline=\nocontentsline#1{#2}\egroup}

\begin{document}

\title{An Optomechanical Platform for Quantum Hypothesis Testing for Collapse Models}

\author{Marta Maria Marchese}
\affiliation{Centre for Theoretical Atomic, Molecular, and Optical Physics, School of Mathematics and Physics, Queens University, Belfast BT7 1NN, United Kingdom}
\author{Alessio Belenchia}
\affiliation{Centre for Theoretical Atomic, Molecular, and Optical Physics, School of Mathematics and Physics, Queens University, Belfast BT7 1NN, United Kingdom}
\author{Stefano Pirandola}
\affiliation{Computer Science and York Centre for Quantum Technologies,University of York, York YO10 5GH, United Kingdom}
\author{Mauro Paternostro}
\affiliation{Centre for Theoretical Atomic, Molecular, and Optical Physics, School of Mathematics and Physics, Queens University, Belfast BT7 1NN, United Kingdom}

\begin{abstract}
Quantum Hypothesis Testing has shown the advantages that quantum resources can offer in the discrimination of competing hypothesis. Here, we apply this framework to optomechanical systems and fundamental physics questions. In particular, we focus on an optomechanical system composed of two cavities employed to perform quantum channel discrimination. We show that input squeezed optical noise, and feasible measurement schemes on the output cavity modes, allow to obtain an advantage with respect to any comparable classical schemes. We apply these results to the discrimination of models of spontaneous collapse of the wavefunction, highlighting the possibilities offered by this scheme for fundamental physics searches.   
\end{abstract}

\maketitle

\section{Introduction}
Hypothesis testing (HT) is an hallmark of any statistical inference toolkit, allowing to discern between the outcomes resulting from the occurrence (or lack thereof) of unknown stochastic processes whose events occur with a set of \textit{a priori} probabilities. Quantum hypothesis testing (QHT), initially introduced for state discrimination tasks~\cite{helstrom1976quantum,chefles2000quantum,barnett2009quantum,weedbrook2012gaussian}, has been applied to channel discrimination and dynamics and its technological potentials in fields like quantum sensing and data read-out are under active investigation~\cite{2020arXiv200410211O,2020arXiv200613250S,2020arXiv201010855H,2020arXiv201003594B,2020SciA....6B.451B}. The characteristic of QHT protocols is that they allow to gain an advantage, in terms of lower error probabilities and in certain parameters range, over \textit{any} classical HT strategy by exploiting quantum resources (like entangled squeezed light).

With the advent of the \textit{second quantum revolution}, quantum technologies manipulating individual quantum systems and employing exquisitely quantum resources to perform tasks are becoming a reality. Crucially, this has also renewed the interest for fundamental investigations of some of the foundational puzzles of quantum theory. Among them, the quantum-to-classical (QtC) transition -- i.e., the process through which the classical world we experience in our daily life emerges from quantum mechanical building blocks~\cite{zurek1991quantum} -- plays a prominent role. Indeed, getting a grasp of the mechanisms governing the QtC could potentially settle some of the interpretational hurdles of quantum mechanics and possibly determine the ultimate limits of validity (if any) of quantum theory itself. 

Collapse models~\cite{RevModPhys.85.471} are one of the most prominent attempts at modifying quantum theory by promoting the collapse of the wavefunction to a physical process embedded in the laws of dynamics by a stochastic modification of the Schr\"{o}dinger equation. In these models, microscopic systems evolves essentially undisturbed by the stochastic collapse, recovering all the predictions of quantum mechanics, while macroscopic objects are subject to a strong localization in position space essentially ruling out Schr\"{o}dinger's cat--like superpositions. One of the most studied collapse models is the so called Continuous Spontaneous Localization model (CSL)~\cite{BASSI2003257}, whose phenomenology has received considerable attention in the last few years~\cite{RevModPhys.85.471,bassi2014collapse,PhysRevLett.125.100404,2020arXiv200109788C,carlesso2019collapse,carlesso2019opto}. In light of this fact, and due to the simplicity of the model, in this work we will focus on CSL. It is important to note that, while CSL postulates a stochastic modification of the Schr\"{o}dinger equation, at the phenomenological level the effect of the model is captured entirely by a dissipative term appearing in the master equation describing close quantum systems dynamics. It is thus clear that, in realistic situations the omnipresence of the environment, and thus the open character of the dynamics, requires sophisticated estimation and inference techniques to discriminate the presence or lack thereof of the CSL mechanism. 

Since collapse models recover all the predictions of quantum mechanics for miscroscopic systems, it is clear that tests able to constrain the parameter space of such models should employ mesoscopic quantum systems. However, creating large spatial superposition of mesoscopic objects is inherently challenging and the subject of intense investigation~\cite{hornberger2012colloquium,arndt2014testing,kaltenbaek2016macroscopic,romero2011large}. Fortunately, collapse models can also be probed via non-interferometric techniques~\cite{Carlesso2018, PhysRevResearch.2.013057,Piscicchia_2017, PhysRevLett.116.090402,vinante2017improved} which do not require the creation and verification of large superpositions. This is exactly the case we explore in this work where the CSL affect the mesoscopic mechanical oscillator in an optomechanical set-up. It is well known that in this situation, the effect of the CSL can be interpreted as an extra mechanical dumping source or, alternatively, as an increased equilibrium temperature of the oscillator. 
{Owing to this, several non-interferometric tests of collapse models have been proposed, and experiments have been carried out constraining the parameter space of the CSL model [cf. Ref.~\cite{RevModPhys.85.471} and references therein]. Inference techniques translated from quantum metrology and estimation theory have been widely employed in such endeavours. Recently,  hypothesis testing has been embedded in theoretical schemes for the assessment of macrorealism and collapse mechanisms~\cite{PhysRevA.100.032111,PhysRevResearch.2.033034}. In this work, we consider hypothesis testing for channel discrimination to probe the dynamical effects of collapse models on macroscopic mechanical oscillators.} 

We thus face the challenge of discriminating between two quantum channels, encoding the presence or lack of CSL, characterizing the dynamics of the mechanical mode. QHT is particularly apt to this task and we set to show that quantum resources can be used to overcome any comparable classical strategy. Furthermore, we propose in the following a specific measurement strategy to preform the hypothesis testing based on realistic parameters for the optomechanical systems. In this way, we do not aim to establish the ultimate advantages that a QHT strategy can allow -- which could result in a hardly feasible measurement scheme -- but we explicitly spell out one strategy that is both feasible within current technology and that presents the aforementioned quantum advantage.   

The remainder of this work is organized as follow. In Sec.~\ref{system} we introduce the optomechanical set-up of interest and we spell out the effect of the CSL on the dynamics of the mechanical mode. In Sec.~\ref{measur} we lay down the measurement schemes that we are going to consider in our analysis. Sec.~\ref{analysis} summarises the main concepts of hypothesis testing and introduces the classical bound we are going to compare the quantum case with. Sec.~\ref{results} presents the main results of our work. Finally, we concluded in Sec.~\ref{conclusions} with a discussion of our findings.

\section{The System}\label{system}
\begin{figure}
    \centering
    \includegraphics[width=0.5\textwidth]{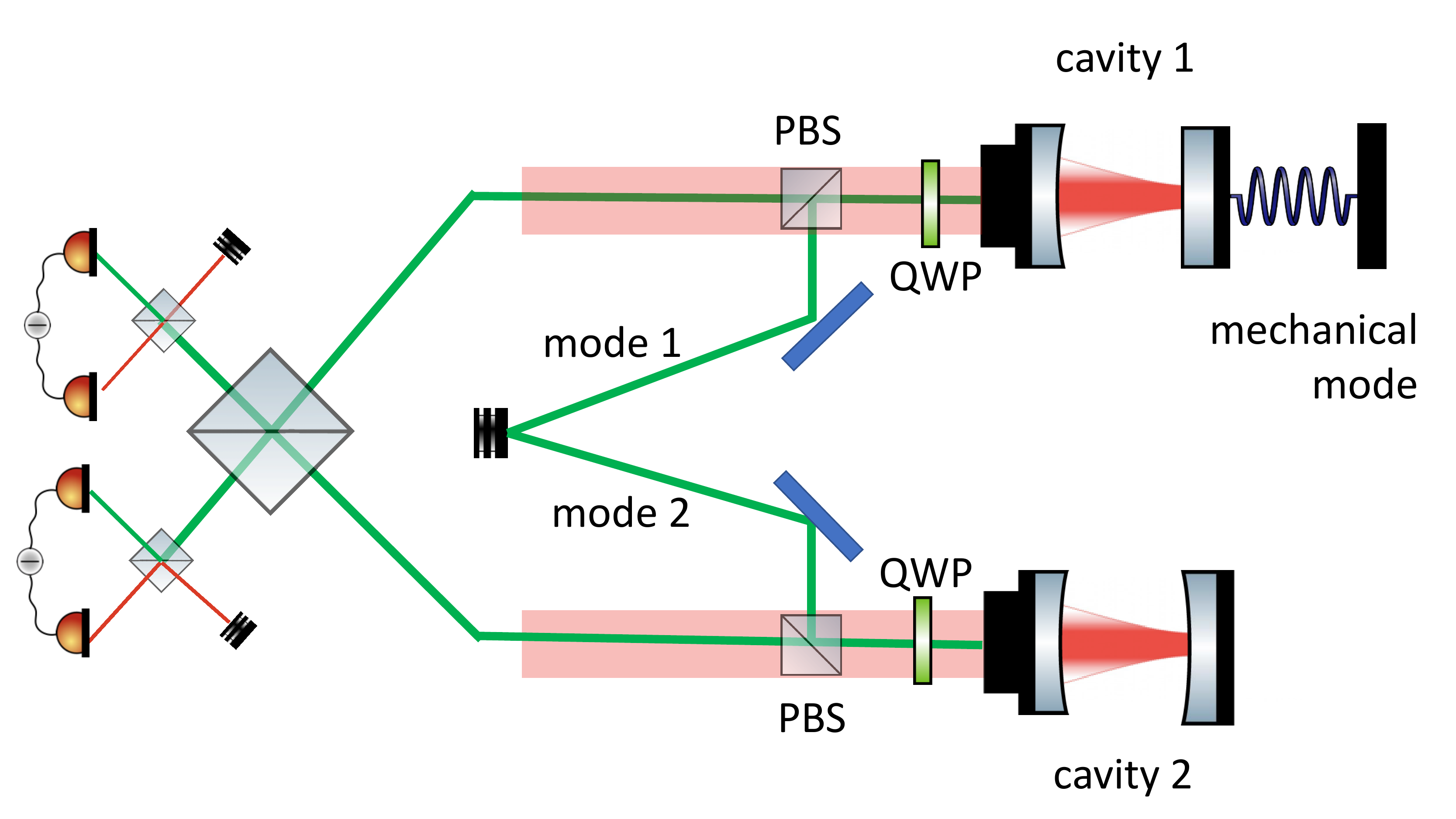}
    \caption{Two cavities, one embedded with a movable mirror, are pumped with a classical laser field (red-shaded region) and with an extra source made of two modes of light (green line). Input modes go through polarizing beam splitters (PBS), enter the cavities, interacting with them and, when they come out, they pass through a quarter-wave plates (QWP), which will change the polarization and will allow us to collect the output modes. Before performing the measurement, we might recombine the outputs using a beam splitter. 
    {The two-cavity set-up is crucial for harnessing the quantum advantage entailed by the use of two-mode squeezed light in a ``quantum reading'' like scheme.}}
    \label{Fig:system}
\end{figure}

Let us consider a system composed by two optical cavities of lenght $L$, as shown in Figure~\ref{Fig:system}. We follow here the discussion in Ref.~\cite{Mazzola2011} where a similar optomechanical system was investigated for entanglement distribution. 
{Note however that here we make use of a single mechanical oscillator, as a second one would not result in a better performance of the scheme proposed in this work.} . The cavity modes with frequency $\omega_C$ are described by creation and annihilation operators $\{\hat{a}^{\dagger}_i,\hat{a}_i\}$, with $i=1,2$. The first cavity is equipped with a movable mirror characterised by position and momentum operators $\{\hat{q},\hat{p}\}$ and damping rate $\gamma_m$. The second cavity is a simple Fabry-Perot cavity with energy decay rate $\kappa$, identical to the first one. They are initially pumped with coherent light with frequency $\omega_L$ and power $P$. The Hamiltonian describing the system reads as

\begin{equation}
\begin{split}
    \hat{H}=&\sum_{i=1,2}\left [\delta \hslash \hat{a}_i^{\dagger}\hat{a}_i +i \epsilon \hslash (\hat{a}_i^{\dagger}-\hat{a}_i)\right]\\
    &+\left(\dfrac{\hat{p}^2}{2m}+\dfrac{m \omega_m^2}{2}\hat{q}^2\right)-\hslash \chi \hat{a}^{\dagger}_1\hat{a}_1\hat{q}
\end{split}
\end{equation}

\noindent
where $\delta=\omega_C-\omega_L$ is the cavity-pump detuning, $\omega_m$ is the frequency of the mechanical oscillator, $\chi={\omega_C}/{L}$ is the radiation-pressure coupling constant and $\epsilon=\sqrt{{2kP}/{\hslash\omega_C}}$ is the amplitude of the laser field which we treat as classical from now on.

When the pumping field is intense enough, as we assume in the following, the description of the dynamics simplifies enormously since we can linearise both the cavity and mechanical modes around their respective steady-state. We thus consider the dynamics of the sole zero-mean quadratures fluctuations that we order in the vector
\begin{equation}\label{ordering}
    \hat{\mathbf{r}}=\left(\hat{Q},\hat{P},\hat{X}_1,\hat{Y}_1,\hat{X}_2,\hat{Y}_2\right)^{\intercal}.
\end{equation}
Here, the first two elements are the dimensionless quadratures for the mechanical mode
\begin{align}
    \hat{Q}=\sqrt{\dfrac{m\omega_m}{\hslash}} \hat{q} \qquad \hat{P}=\dfrac{1}{\sqrt{\hslash m \omega_m}}\hat{p}
\end{align}
with $\big[\hat{Q},\hat{P}\big]=i$. The remaining components of the vector represent the optical quadratures
\begin{align}
    \hat{X}_j=\dfrac{\hat{a}_j^{\dagger}+\hat{a}_j}{\sqrt{2}}, \qquad
    \hat{Y}_j=i\dfrac{\hat{a}_j^{\dagger}-\hat{a}_j}{\sqrt{2}} \qquad (\text{with}\, i=1,2),
\end{align}
for the two intra-cavity field modes.

The quadrature vector evolves in time according to the Langevin equations in the input-output formalism
\begin{equation}
    \dot{\hat{\mathbf{r}}}=A\hat{\mathbf{r}}+\hat{\mathbf{n}}.
\end{equation}
where the $6\times 6$ drift matrix $A$ is given by
\begin{equation}
    A=\begin{pmatrix}
    0       &       \omega_m        &       0        &      0       &       0       &       0\\
    -\omega_m   &   -\gamma_m & \sqrt{2}\alpha g    &       0       &       0       &       0\\
    0       &       0        &       -\kappa           &     \delta &   0   &   0\\
    \sqrt{2}\alpha g    &   0   &   -\delta    &   -\kappa      &       0   &   0\\
    0       &       0       &       0       &       0       &       -\kappa      &        \delta\\    0       &       0       &       0       &       0       &       -\delta      &    -\kappa\\
    \end{pmatrix}\label{driftmatrix}
\end{equation}
\noindent
with $\alpha= \Re[\langle a\rangle]$ the square root of the number of photons in the cavity and $g=\chi\sqrt{{\hslash}/{m\omega_m}}$ the effective coupling rate. The vector $\hat{\mathbf{n}}$ collects the zero-mean quantum noise operators and it is given by
\begin{equation}
\hat{\mathbf{n}}=\left(0, \hat{\xi}, \sqrt{2\kappa}\hat{X}_{in_1},\sqrt{2\kappa}\hat{Y}_{in_1}, \sqrt{2\kappa}\hat{X}_{in_2},\sqrt{2\kappa}\hat{Y}_{in_2}\right)^{\intercal}.\label{Eq:noisevector}
\end{equation}
Here, $\hat{\xi}$ is a Langevin force operator encoding the interaction of the mechanical mode with a phononic thermal bath at temperature $T$ and producing the Brownian motion of the mechanical oscillator. This noise is characterised by its two-point correlator which can be written as 
\begin{equation}
    \langle \xi(t)\xi(t')\rangle=\dfrac{2\gamma_m k_B T}{\omega_m\hbar}\delta(t-t'),
\end{equation}
in the high temperature limit $k_B T\gg\hbar\omega$~\cite{PhysRevA.63.023812}.
\noindent
Then $\left\{\hat{X}_{in_k},\hat{Y}_{in_k}\right\}$, with $k=\{1,2\}$, are the quadratures of the input noises impinging on the two cavities.
The covariance matrix of these two input modes encodes the information on the light state we feed to the cavities on top of the coherent pumping.

In the linearized picture that we are considering, the total Hamiltonian of the system is at most quadratic in the quadratures $\hat{\mathbf{r}}$ 
{while the Lindblad operators, describing the interaction with the phononic thermal bath, are at most linear in them.} Thus, if both the initial state of the modes $\hat{\mathbf{r}}$ and the state of the input noises are Gaussian, the dynamics will preserve this Gaussianity~\cite{Ferraro2005a,genoni2016conditional}. This observation enormously simplify the dynamics of the system since it is enough  to consider the evolution of the first and second statistical moments of the quadratures of the system. Furthermore, as we consider zero-mean quantum fluctuations 
{and the dynamics of the mean values is decoupled from the evolution of the variances}, it is sufficient to work with the time evolution of the covariance matrix 
{$\bm{\sigma}$}
, which is ruled by the Lyapunov-like equation
 \begin{equation}
     \dot{\bm{\sigma}}= A \bm{\sigma} +\bm{\sigma} A^{T} +D,\label{Eq:equazionemoto}
 \end{equation}
where $D$ is the so-called diffusion matrix. The elements of $D$ depend on the two-point correlations of the noise vector as~\cite{Mari2009}
\begin{equation}
    D_{ij}=\dfrac{1}{2}\left[ \langle n'_i(t)n'_j(t)\rangle+\langle n'_j(t)n'_i(t)\rangle\right].
\end{equation}
Considering the aforementioned sources of noise, we can express the $6\times 6$ diffusion matrix $D$ in block-diagonal form as 
\begin{equation}\label{Dmatrix}
     D=\begin{pmatrix}
     \bm{\sigma}_m & 0\\
     0 & \bm{\sigma}_{IN}
     \end{pmatrix}
\end{equation}
where  $\bm{\sigma}_{IN}$ is the 4$\times$4 dimensionless covariance matrix associated to the input modes times $2\kappa$, while 
\begin{equation}
\bm{\sigma}_m=
\begin{pmatrix}
 0 & 0\\
 0 & 2 \dfrac{\gamma_m k_B T}{\hbar \omega_m} + \Delta
\end{pmatrix}\label{Eq:sigmam}
\end{equation}
is the 2$\times$2 matrix describing the thermal dissipation. 

Note that, in the last term entering the diffusion matrix we have introduced an extra 
{heating rate} parameter $\Delta$. This parameter \textit{de facto} modifies the equilibrium temperature of the mechanical oscillator and correspond to an extra dissipation channel for the open quantum system composed by the two cavity modes and the mechanical one. Before moving on, let us briefly remark in the following that the stochastic effect of the CSL model on the mechanical mode in cavity one is encoded exactly in the extra parameter $\Delta$ appearing in the diffusion matrix.

\subsection{Continuous Spontaneous Localisation model}
Collapse models (CMs)~\cite{RevModPhys.85.471} introduce stochastic modifications to the Schr\"odinger equation of quantum mechanics in the attempt to promote the collapse of the wavefunction to a dynamical process providing a dynamical picture of how the classical world emerges from the quantum microscopic one. For our purposes here we do not need to go into the details of collapse models. It is enough to say that we will make use of the arguably better studied among CMs, the so called Continuous Spontaneous Localization model (CSL). 

The CSL, with white noise, 
{describes the collapse as a continuous process in time. This introduces in the master equation of the system an extra spatial decoherence term whose phenomenology} is completely characterized by two parameters $\{r_C,\gamma\}$. The parameter $r_C$, is 
{the localization length of the model, i.e., }the characteristic length-scale above which the collapse mechanism is relevant. 
{The collapse rate $\gamma$ sets the strength of the CSL  mechanism~\cite{RevModPhys.85.471}.}

In our setting, the CSL mechanism affects in a significant way only the mechanical mode due to its ``mesoscopic nature'' in view of the fact that CMs are formulated in such a way that their predictions deviate from standard quantum mechanics only for meso-/macroscopic systems. Our mechanical mode is in contact with a thermal phonon bath at temperature $T$, whose effect is described by the operator $\hat{\xi}$ in equation \eqref{Eq:noisevector}. On top of that, we consider the decoherence induced by the CSL. Formally, we can treat the effect of the CSL by defining a modified equilibrium temperature of the oscillator via
\begin{equation}
    \gamma_m (2\bar{n}_{th}+1)+\Delta=\Delta(2n_{CSL}+1)\rightarrow n_{CSL}=\bar{n}_{th}+\frac{\Delta}{2\gamma_m}.\label{CLStemperature}
\end{equation}
where $\bar{n}_{th}$ is the thermal number of phonons at temperature $T$. The parameter $\Delta$ entering this expression and the diffusion matrix in~\eqref{Eq:sigmam} is a function of both $r_c$ and $\gamma$ as well as the mass distribution of the system of interest. As reported in~\cite{PhysRevLett.112.210404} it can be written explicitly as 
\begin{equation}\label{csl_rate}
    \Delta=\frac{\hbar\gamma}{3m\omega_m m_0^2}\sum_{k=1}^3\int \frac{e^{-\frac{|\mathbf{r}-\mathbf{r}'|^2}{4r_C^2}}}{(2\sqrt{\pi}r_C)^3}\partial_{r_k}\rho(\mathbf{r})\partial_{r'_k}\rho(\mathbf{r}'){\rm d}\mathbf{r}{\rm d}\mathbf{r}',
\end{equation}
where $m_0=1$~amu (atomic mass unit) and $\rho(\mathbf{r})$ is the mass density of the system subject to the CSL.

\section{Measurement schemes}\label{measur}
The main objective of this work is to investigate the potential of optomechanical systems and quantum reading protocols~\cite{Pirandola2011,PhysRevLett.106.090504} for the discrimination of the additional dissipative channel described by the parameter $\Delta$ via the methods of quantum hypothesis testing. In particular, we want to determine whether quantum resources, in the form of non-classical input noise states, can lead to a quantum advantage with respect to classical resources for hypothesis testing. In order to accomplish this, we will examine two cases for the state of the input noise modes: (i) a two-mode squeezed state as an entangled, and thus quantum, resource and (ii) two independent thermal states as classical ones. 

\begin{figure}
    \centering
    \begin{minipage}[t]{0.45\linewidth}
    \includegraphics[width=1\textwidth]{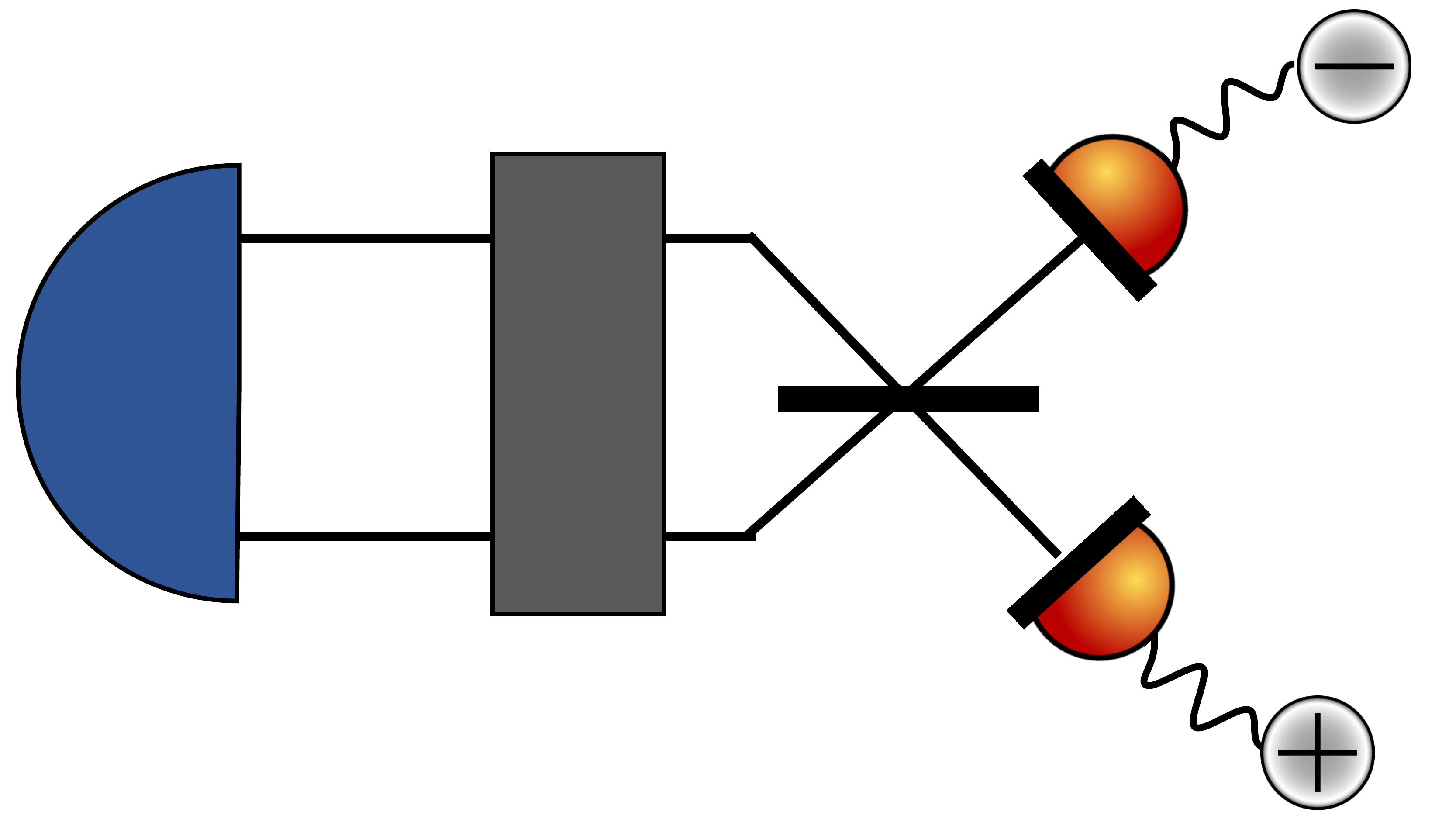}
    \end{minipage}
    \hfill\vline\hfill
    \begin{minipage}[t]{0.45\linewidth}
    \includegraphics[width=1\textwidth]{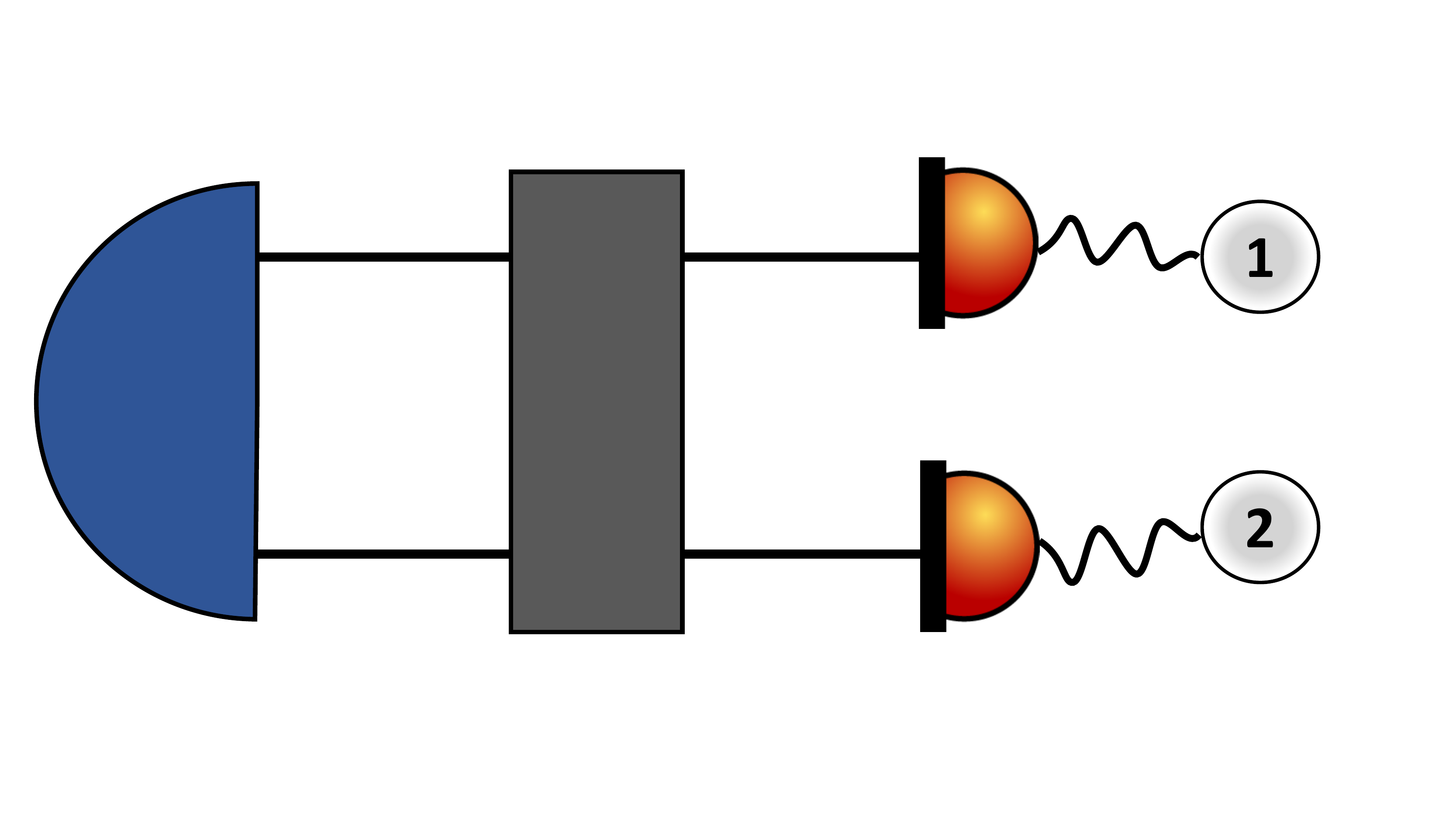}
    \end{minipage}
       \caption{On the left a scheme for the measurement of EPR quadratures, while on the right for a local measurement. The scheme in Fig.~\ref{Fig:system}, captures the case on the left here, i.e., EPR quadratures measurement. In order to implement the classical scheme on the right, it would be sufficient to remove the beam-splitter in Fig.~\ref{Fig:system} combining the output cavity modes.}
       \label{Fig:measurements}
\end{figure}

Furthermore, in order to probe the system the output modes emerging from the two cavities needs to be measured. Also at this stage we can consider different measurement strategies with different ``degrees of quantumness'' at play, see Fig.~\ref{Fig:measurements}. The output modes can be directed towards photodetectors by using a combination of quarter waveplates and polarised beamsplitters (see Fig.~\ref{Fig:system}). We then consider two different measurement schemes: (i) a local measurement, consisting in measuring directly the quadratures of the output modes $\{\hat{x}_{out_i},\hat{y}_{out_i}\}$ with $i=1,2$; (ii) in the spirit of the original quantum reading protocol~\cite{Pirandola2011}, the output modes can be further recombined through another beamsplitter to perform a measurements of EPR-like quadratures $\{\hat{q}_{\mp},\hat{p}_{\pm}\}$ of the emerging modes $\{+,-\}$. These are defined as
\begin{equation}
\begin{cases}
\hat{q}_{\mp}=\dfrac{\hat{x}_{out_1}\mp\hat{x}_{out_2}}{\sqrt{2}}\\
\hat{p}_{\pm}=\dfrac{\hat{y}_{out_1}\pm\hat{y}_{out_2}}{\sqrt{2}},
\end{cases}
\end{equation}
in term of the output modes.

Finally note that, using the input-output formalism~\cite{WallsMilburn}, the output modes can be easily expressed in terms of the input ones via
\begin{equation}
\begin{cases}
     \hat{x}_{out_{1,2}}=&\sqrt{\kappa}\hat{X}_{1,2}-\hat{X}_{in_{1,2}} \\
    \hat{y}_{out_{1,2}}=&\sqrt{\kappa}\hat{Y}_{1,2}-\hat{Y}_{in_{1,2}}.  
\end{cases}
\end{equation}
All the output modes will depend on the parameter $\Delta$, which vanishes in the case where no CSL is present. For simplicity of notation we omit the dependence on this parameter. 
{It should also be noticed that, such dependence arises owing to the dynamics of the mechanical system. By itself, the CSL mechanism does not influence light and this guarantee the read-out of the CSL effect on the mechanical oscillator in our set-up}.

\subsection{Initial state}\label{SubSec:initialstate}
While not strictly part of the measurement strategy, we comment here on the initial state of the system (cavities+mechanical mode) that we will assume in the rest of the work unless otherwise stated. Indeed, the choice of a particular initial state is undoubtedly part of any  protocol and one on which the feasibility of the protocol hinge. 

In view of these considerations, we consider as initial state the product state of the cavities' steady-states when subject to vacuum input noises (i.e., when only the coherent pumping is present). The result is a state in which the mechanical mode and the first cavity reach their joint steady-state
{, fully characterized by the covariance matrix} $\bm{\sigma}_{ss}^{\rm{mc}}$, while the optical mode of the second cavity remains in its ground state
\begin{equation}
\bm{\sigma}_{ss}=\left(\begin{array}{c|c}
\bm{\sigma}_{ss}^{\rm{mc}}  & \mathbb{O}_{4\times 2} \\ \hline 
\mathbb{O}_{2\times 4}  & \mathbb{I}_{2\times 2}/2
\end{array}\right),
\end{equation}
where $\mathbb{I}_{n\times m}$ and $\mathbb{O}_{n\times m}$ are $n\times m$ identity and zeros matrices, respectively.
This choice of initial state corresponds in practice to commencing the experiment without additional light sources and wait long enough for the system to reach its steady-state. After this, we can start to probe the cavities with extra modes of light (the input noises) and measure the output cavity fields as described above. 

\subsection{Protocol description}
Before proceeding further and introducing the hypothesis testing, let us summarize the steps of our channel discrimination protocol:
\begin{enumerate}
    \item \textbf{Preparation:} the optomechanical system, subject to only the coherent pumping of the cavities (i.e., vacuum input noise) reaches its steady-state;
    \item \textbf{Classical/quantum resources:} two additional light mode impinge on the cavity and are prepared in either a two-mode squeezed state (TMS) or in the tensor product of two local thermal states. The system starts to evolve away from the initial state;
    \item \textbf{Measurements:} at the same time as the input noises are fed into the cavity, we can monitor the output modes via photodetectors. This can be perform either via local measurement or measuring EPR-like quadratures after recombining the output modes via a beamsplitter;
    \item \textbf{Post-processing:} the obtained measurement outputs are post-processed via a $\chi^2$-test to discriminate the possible channels, i.e. if the parameter $\Delta$ vanishes or not.
\end{enumerate}
While we wil discuss in detail the hypothesis testing post-processing in the next section, it should be noted that the $\chi^2$-test is suitable since the output of the quadratures measurements follow a Gaussian distribution with zero-mean and variance depending on $\Delta$.

\section{Quantum Hypothesis Testing}\label{analysis}
In this section, we summarize the main elements of hypothesis testing and specify the classical bound to the error probability in our specific set-up. This lays the basis for comparison between classical and quantum protocols and to show an advantage in using quantum resources. 

In a typical binary hypothesis testing, two exclusive hypotheses are  formulated. Hypothesis $H_0$ is called null hypothesis and it is the starting point: we assume this to be true and we will conduct a test to determine whether this is likely to hold or not. The alternative hypothesis $H_1$ contradicts the previous one and expresses what we think is wrong in the null hypothesis. In our set-up, we aim at testing whether the dissipative channel associated to $\Delta$ is present. This also means testing if the effect coming from CSL model is present or not. In this context, $H_0$ corresponds to no new physics, i.e. an open dynamics with no CSL, while $H_1$ to the presence of the extra dissipative mechanism. 

The hypothesis testing is performed by post-processing the measurement outcomes. As highlighted in the previous section, these outcomes follow zero-mean Gaussian distributions. This implies that the hypothesis testing, and so the channel discrimination, corresponding to discriminating two Gaussians with different variances 
{($V_0,\, V_1$)} depending on $\Delta$. Thus we can formulate the two hypotheses as follow
\begin{equation}
\begin{cases}
H_0:\Delta=0 \iff  V=V_0\\
H_1:\Delta>0 \iff  V=V_1 \neq V_0.
\end{cases}
\end{equation}
Moreover, it is easy to verify that, for $\Delta>0$ the condition $V_1 > V_0$ holds for both the outcomes of local measurement of the output variables and EPR-like variables. This implies that we can conduct a one-tail test.

{It is important to note that, in general,  $\Delta\geq 0$, with $\Delta=0$ corresponding to the absence of the CSL. Therefore, statistical inference methods can only rule out, with a certain likelihood, some parts of the CSL parameter space casting upper bounds on $r_C$ and $\gamma$.}

Given the nature of the problem, we use a $\chi^2$-test in the following by defining the test statistic $T=(N-1)s^2/{V_0}$, where $s^2=\sum_{i=1}^{N}{(r_i-\bar{r}_i)^2}/{N-1}$ is the sample variance for a sample-size $N$, and $r_i$ is the variable we decide to use for the test among $\{q_\pm ,p_\mp\}$, for EPR-like measurements, or $\{x_{out_{1,2}}, y_{out_{1,2}}\}$ for the classical ones. The test statistic follows a $\chi^2$-distribution with $N-1$ degrees of freedom. Note that, contrary to the quantum reading protocol in~\cite{Pirandola2011}, in our case for each measurement schemes different quadratures have different variances. This means that they should be subjected to separate tests not allowing to double the number of outcomes as in~\cite{Pirandola2011}. 

In an experiment, the hypothesis testing proceed by comparing the likelihood for the particular realization of the test statistic $T=t^*$ with the so called \textit{significance level} $\alpha$ of the test, i.e., the maximum error that we allow ourselves to commit by rejecting $H_0$ when true. In particular,
\begin{equation}
\begin{cases}
\text{If }\; t^{*}\leq Q_{1-\alpha}^{N-1}  \Leftrightarrow P|_{H_0}(T\geq t^{*})\leq \alpha \; \Rightarrow \;\text{Reject $H_0$}\\\\
\text{If }\; t^{*}<Q_{1-\alpha}^{N-1}  \Leftrightarrow P|_{H_0}(T\geq t^{*})> \alpha \; \Rightarrow \;\text{Accept $H_0$}.
\end{cases}
\end{equation}
Here we indicate with $P|_{H_0}( T\geq t^{*})$ the probability of obtaining a value of the random variable $T$ larger then $t^*$ conditioned on assuming $H_0$ is true. Note that, as usual, the condition on the probability are mirrored by condition on the realization of the statistic in terms of the quantiles of the $\chi^2$-distribution $Q_{1-\alpha}^{N-1}$. 

Crucial quantities in hypothesis testing are the error probabilities, i.e. the probability of rejecting $H_0$ when true and the probability of accepting $H_0$ when false. The former is known as Type I error and quantified by $P(H_1|H_0)$, the latter is a Type II error quantified by $P(H_0|H_1)$. Assuming error priors for the two hypotheses, the mean error probability is $P_{err}=\left(P(H_1|H_0)+P(H_0|H_1)\right)/{2}$.
It is a simple exercise to find the expression for the total error probability given by 
\begin{equation}
   P_{err}= \frac{1}{2} \left[1-\frac{\Gamma \left(\frac{N-1}{2},\frac{Q_{1-\alpha}^{N-1} V_0}{2 V_1}\right)}{\Gamma
   \left(\frac{N-1}{2}\right)}+\frac{\Gamma \left(\frac{N-1}{2},\frac{Q_{1-\alpha}^{N-1}}{2}\right)}{\Gamma \left(\frac{N-1}{2}\right)}\right],
\end{equation}
where $\Gamma(z,x)$ and $\Gamma(z)$ are the incomplete and complete Gamma functions, respectively.

Finally, while the values of $V_{0,1}$ entering the error probability expression depend on both the initial noise state and the measurement scheme, their functional form depends only on the latter. In particular, using the input-output relations and the ordering of the system degrees of freedom for the elements of covariance matrix of the system $\bm{\sigma}(t)$ as in Eq.~\eqref{ordering} it is easy to obtain the expressions for the variances of the output results. For example we have
\begin{align}
    & {\rm Var}(x_{out,1})=2\kappa\sigma_{33}\\
    & {\rm Var}(q_\pm)=\kappa\left(\sigma_{33}+\sigma_{55}\pm 2\sigma_{35}\right),
\end{align}
{where $\sigma_{ij}$ are the elements of the covariance matrix solving Eq.~\eqref{Eq:equazionemoto},}and analogously for the rest of the measured quadratures. The dependence of these expressions on the initial noises, as well as on the unknown parameter $\Delta$, is hidden in the elements of the covariance matrix $\bm{\sigma}(t)$ coming from solving the dynamics. We identify $V_{0,1}$ in the hypothesis testing with the values of the relevant variances -- depending on the measurement scheme and initial noise chosen -- for $\Delta=0$ or $\Delta>0$ respectively.

\subsection{Classical bound}\label{SubSec:classicalbound}
We now show that, at intermediate times, quantum resources allow to attain a total error probability lower than the one achievable by any comparable classical strategy. In order to claim this, we need a measure of the minimum error probability attainable. Following the results in \textit{theorem 4} of~\cite{Pirandola2011} for the discrimination of two Gaussian channel via a classical protocol, the error probability is lower-bounded by
\begin{equation}
    {\cal C}(n_1,n_2,t):=\dfrac{1- \sqrt{1- (F(n_1,n_2,t))^N}}{2}
    \label{C_bound}
\end{equation}
where $F(n_1,n_2,t)$ is the fidelity between the two-mode output Gaussian states corresponding to the evolution of the system up to time $t$ when $\Delta=0$ or $\Delta>0$ with classical input noise thermal states characterized by $n_1$ and $n_2$ mean photon numbers ~\cite{Banchi2015}\footnote{It should be noted that, what we call here fidelity ($F$) corresponds to the fidelity squared ($\mathcal{F}^2$) in~\cite{Banchi2015}}. $N$ is again the number of measurement outcomes collected at time $t$. 

{Eq.~\eqref{C_bound} is the most stringent bound to the error probability to discriminate the two Gaussian states $\rho_{\Lambda=0}$ and $\rho_{\Lambda\neq 0}$~\cite{Banchi2015}. It is obtained from the form for the error probability derived by Helstrom~\cite{helstrom1976quantum}, $P_{err}(\rho_0,\rho_1)=1-D(\rho_0,\rho_1)$ where $D(\rho_0,\rho_1)$ is the trace distance, by using the inequality $D(\rho_0,\rho_1)\leq\sqrt{1-F(\rho_0,\rho_1)}$ and the factorization properties of the quantum fidelity for product states -- $\rho(t)=\bigotimes_{k=1}^N \rho_k(t)$, with $\rho_k$ the two-mode Gaussian state in each run of the experiment at time $t$.}

\section{Results}\label{results}

We aim to show that, the total error probability for the channel discrimination in our set-up, when the input noises are quantum correlated, can be lower than the one that can be achieved by \textit{any} comparable classical strategy. It should be stressed that we do not aim to find the ultimate quantum bound to the error probability. Indeed, our scope is more practical: we want to show that such an advantage exists for the specific protocol we consider. 
\begin{table}[b]
\centering
\begin{tabular}{llll}
    \hline
    \textbf{Symbol}  & & \textbf{Name} & \textbf{Value}  \\ 
    &&&{\bf or Expression}\\
    \hline\hline
    $\gamma_m$ & & Mechanical dumping & $2\pi \omega_m/10^5$ \\
    $\omega_m/2\pi$ & & Mechanical frequency & $2.75\times 10^5$~Hz \\
    $T$ & & Phononic bath's temperature  & $10^{-3}$~K \\
    $\omega_c/2\pi$ & & Cavity mode's frequency & $9.4\times 10^5 c$ \\
    $m$ & & Mass  & $150$~ng \\ 
    $L$ & & Cavity length & 25~mm \\
    $\kappa$ & & Cavities linewidth & $5\times 10^7$~Hz \\
    $\delta$ & & Cavity-pump detuning & $4\kappa$ \\
    $P$ & & Pump laser power & $4\times 10^{-3}$~W \\
    $R$ & & Mechanical system linear dimension & 1~$\mu$m \\
    \hline
\end{tabular}
\caption{Specifics of all the parameters for the set-up of two cavities and a mechanical mode entering the simulations in this work.}
\label{tab:parameters}
\end{table}

In what follows we fix the significance level to be $\alpha=5\%$ unless otherwise stated. All the values of the system parameters used in the simulations are reported in table~\ref{tab:parameters}.
These values are within reach of current technology which is in favour of the feasibility of the protocol. Finally, we assume $\Delta=10^6$~Hz unless otherwise specified. This value of the parameter, characterizing the unknown extra-channel whose presence we want to discriminate, is such that the extra diffusion associated with it is greater than the thermal diffusion characterized by $2\gamma_m k_B T/(\hslash \omega_m)$ and it is motivated by the CSL model. As we discussed previously, the CSL model with white-noise is completely characterised by the two parameters $r_C$, and $\gamma$. The first, $r_C$ can be fixed at $100$~nm~\cite{Carlesso2018} while for the second one we consider the value proposed by Adler~\cite{Adler_2007} $\gamma_{A}=10^{-28}$~m$^3$Hz. Indeed, assuming the mechanical mode to describe the center of mass of a system with linear dimension $R$, that we approximate as spherical for simplicity, and using Eq.~\eqref{csl_rate}, this choice of CSL parameters results in $\Delta\approx 10^6$~Hz.

We perform a dynamical analysis, starting from the steady-state of our three-partite system when vacuum input noises are present, and focus on the transient before the system reaches a new steady-state. In doing this, we compare two protocols: a classical one using input thermal noises and local measurements of the output modes, and a quantum one with two-mode squeezed input noises and EPR-like measurements. 

In order to show the advantage coming from using quantum resources in our context, we need to compare the quantum scheme with the classical lower bound to the error probability. A fair comparison can be achieved by fixing the photon number in the input noises to the two cavities $\{n_1,n_2\}$ and comparing situations with the same sample size $N$, i.e., repetitions of the experiment. We thus compare the error probability coming from the quantum protocol using a TMS input noise with the lower bound that can be achieved starting from uncorrelated thermal noises with the same mean photon number per input mode as the TMS. The lower bound, as already discussed, is the minimum error probability that can be achieved by any classical measurement procedure  ~\cite{Pirandola2008}. 
Thus, the comparison with it can show possible quantum advantages.  

\begin{figure}
\centering
{\includegraphics[width=1.\columnwidth]{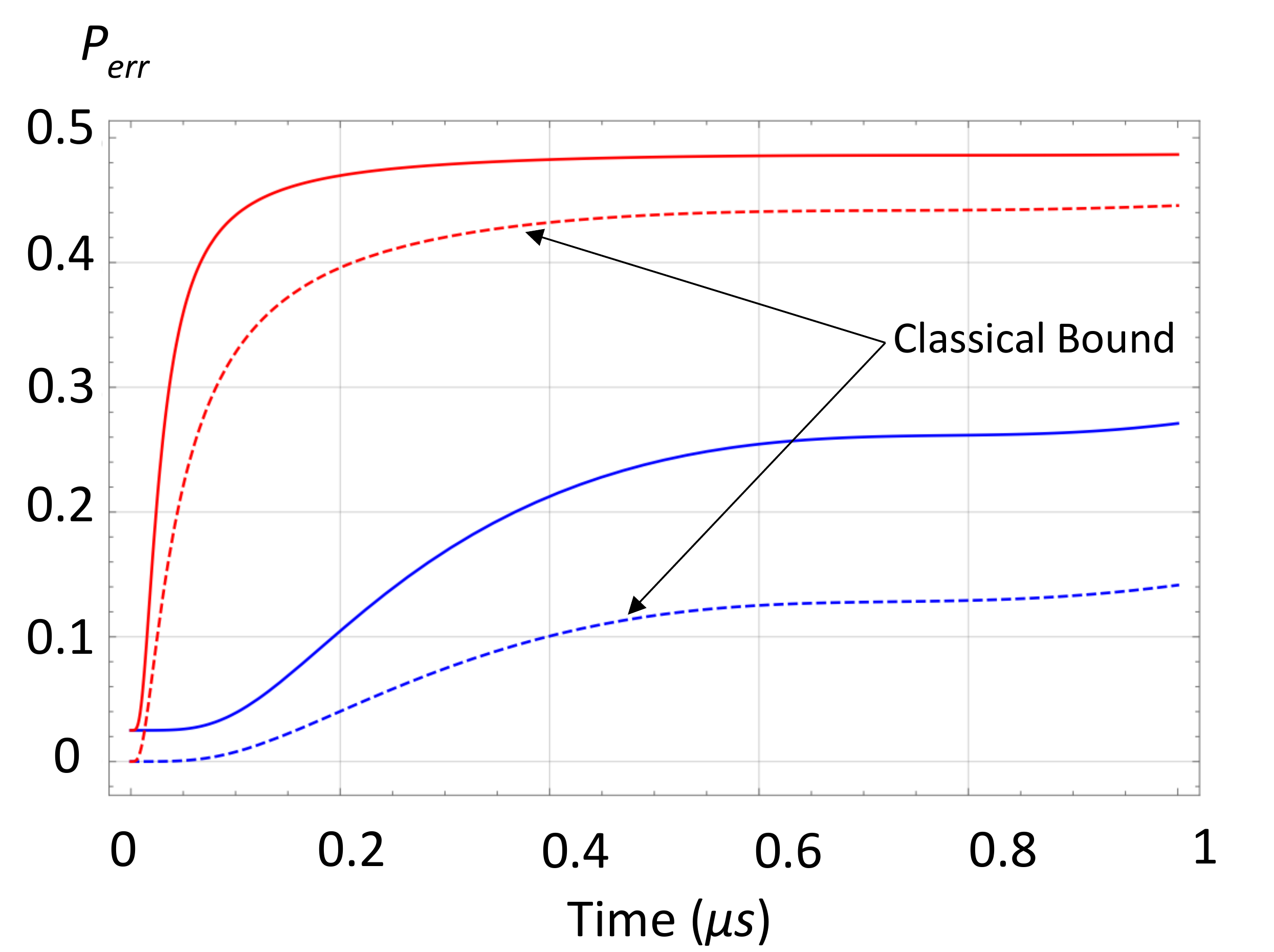}}
\caption{Comparison between the classical error probabilities (solid lines) and the respective bounds $\mathcal{C}$ (dashed lines) for different values of the mean number of photons in the input noise's modes, $n_1=n_2$. The red curves represent $n_1=10$ while the blue ones $n_1=100$ and we consider the statistics of measurements for the $x_{out_1}$ output quadrature. As discussed in the main text, $\Delta=10^6$~Hz which corresponds to CSL with Adler parameters~\cite{Adler_2007}. The level of significance and the number of experiments are $\alpha=5\%$ and $N=100$ respectively.}
\label{fig:Perr_C}
\end{figure}

We start by showing the discrepancy between the local measurement strategy, with classical input noises -- a.k.a. the classical protocol -- with the lower bound to the error probability $\mathcal{C}$. The classical input noise is characterized by its thermal covariance matrix 
\begin{equation}
\bm{\sigma}_{IN}=2\kappa\left(\begin{array}{c|c}
(n_1+1/2)\mathbb{I}_{2\times 2}  & \mathbb{O}_{2\times 2} \\ \hline 
\mathbb{O}_{2\times 2}  & (n_2+1/2)\mathbb{I}_{2\times 2}
\end{array}\right)
\end{equation}
entering Eq.~\eqref{Dmatrix}. Fig.~\ref{fig:Perr_C} shows the classical error probability and the correspondent bound for two values of $n_1$. The value of $n_2$ does not have any bearing on these probabilities. We can see that the classical error probability is always greater than the classical bound, as expected, and the bigger the number of photons we inject as noise in the first cavity the worse our ability to discriminate the two hypotheses become. 

\begin{figure}
\centering
{\includegraphics[width=1.\columnwidth]{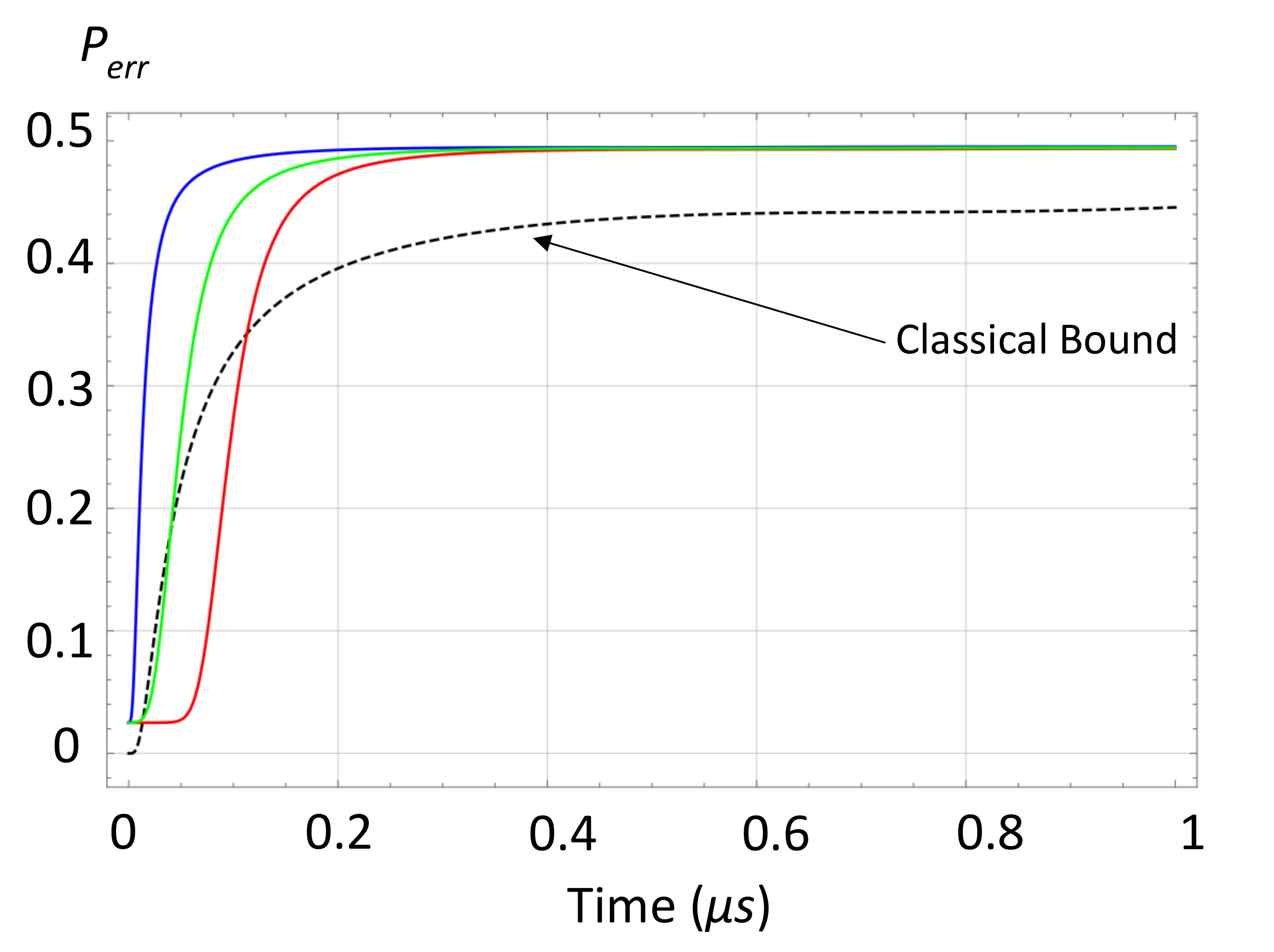}}
\caption{Quantum error probabilities (solid lines) for different squeezing angles $\phi=\{\pi/2,5\pi/6,\pi\}$. The parameters are $n_1=n_2=100$, $N=100$, $\Delta=10^6$~Hz, and $\alpha=5\%$, and we consider the statistic of measurements for the $q_{+}$ output quadrature. The dashed curve represents the corresponding classical bound $\mathcal{C}(n_1,n_2,t)$, the blue curve is the quantum error probability when the squeezing angle is $\phi=\pi/2$, the green one for $\phi=5\pi/6$ and the red one corresponds to $\phi=\pi$. We observe that, for the squeezing angle approaching $\phi=\pi$, violations of the classical bound are possible at intermediate times.}
\label{fig:angolo}
\end{figure}
\begin{figure}
\centering
{\includegraphics[width=1.\columnwidth]{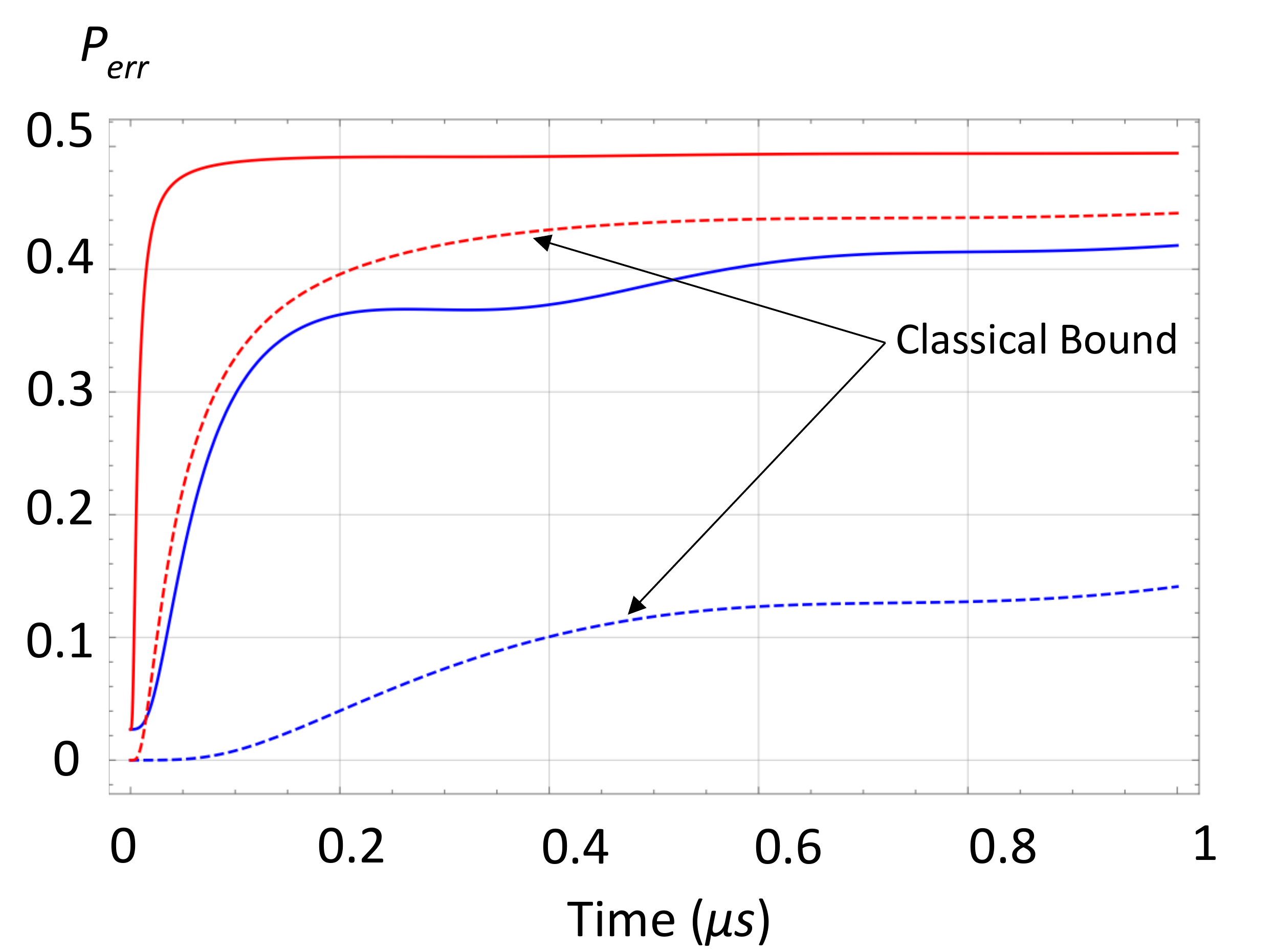}}
\caption{Comparison between quantum error probabilities (solid lines) and the corresponding classical bounds (dashed lines) for vanishing squeezing angle, $\phi=0$. The blue curves are computed for $n_1=n_2=10$ while the red ones for $n_1=n_2=100$. In both cases $N=100$, $\Delta=10^6$~Hz, and $\alpha=5\%$, and we consider the statistic of measurements for the $q_{+}$ output quadrature.}
\label{fig:C_bound_n1}
\end{figure}
\begin{figure}
\centering
{\includegraphics[width=1.\columnwidth]{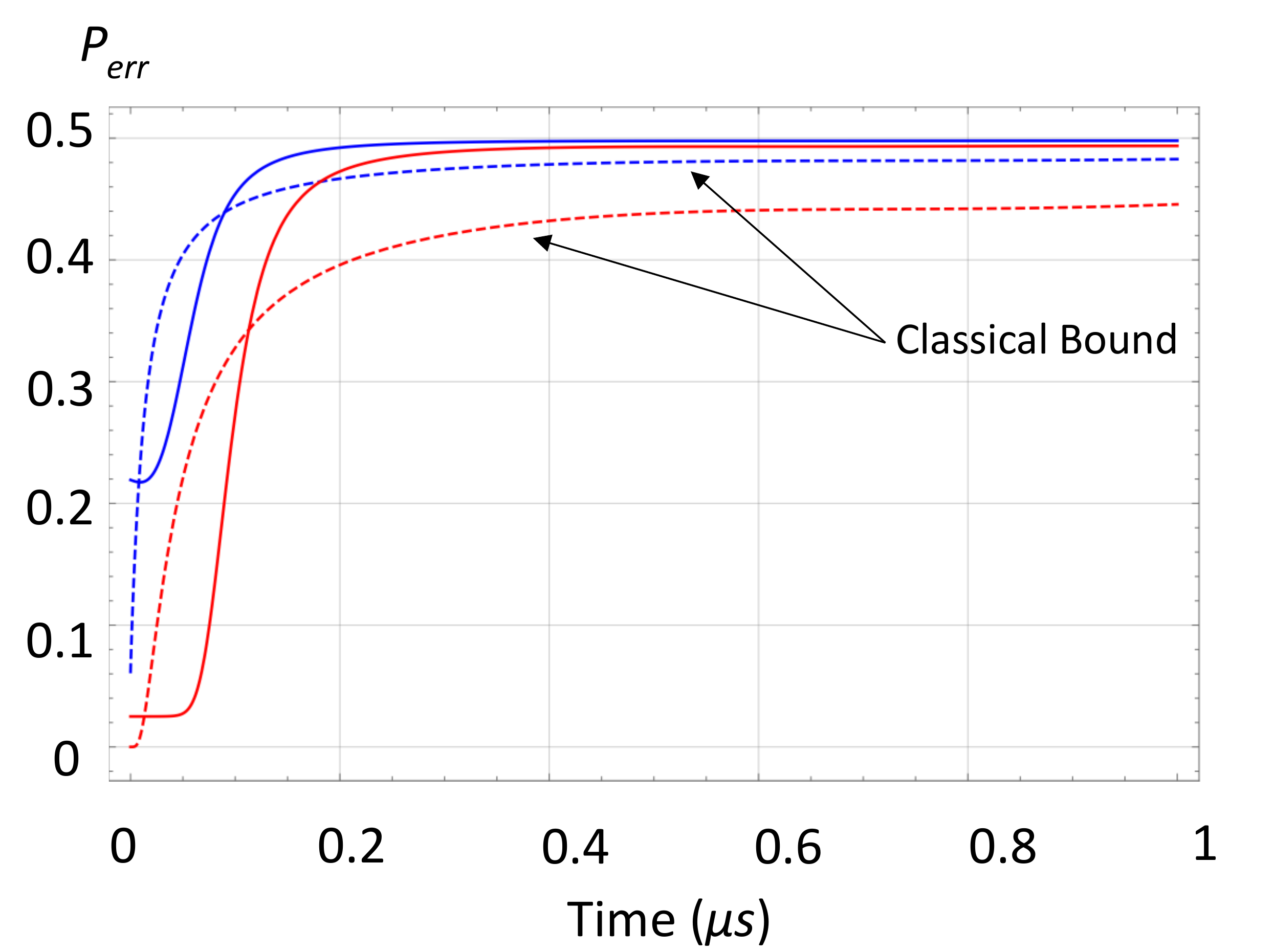}}
\caption{Quantum error probabilities for $\phi=\pi$ (solid lines) and corresponding classical bounds (dashed lines) for varying number of repetitions of the experiment $N={10,100}$. The blue curve corresponds to $N=10$, the red one to $N=100$. We fix $\alpha=5\%$, $\Delta=10^6$~Hz, and $n_1=n_2=100$. We consider the statistic of measurements for the $q_{+}$ output quadrature.}
\label{fig:C_bound_N}
\end{figure}

In Fig.~\ref{fig:angolo}, we show a first comparison of the quantum error probability with respect to the classical bound. In this case, the input noise state is a TMS state with same mean photon number on the two modes and characterised by the squeezing parameter $r\geq 0$ and the squeezing angle $\phi$. The input covariance matrix entering in Eq.~\eqref{Dmatrix} is given by 
\begin{equation}
\bm{\sigma}_{IN}=\kappa\left(
\begin{array}{c|c}
 \cosh 2r \mathbb{I}_{2\times 2}  & \sinh 2r R_{\phi}\\
 \hline
  \sinh 2r R_{\phi}& \cosh 2r \mathbb{I}_{2\times 2}
\end{array}
\right),
\end{equation}
where 
\begin{equation}
    R_\phi=\begin{pmatrix}
     \cos\phi & \sin\phi\\
     \sin\phi & -\cos\phi
    \end{pmatrix}.
\end{equation}
Fixing the mean photon numbers in the two input modes corresponds to fixing the value of the squeezing parameter $r$ given the relation $\cosh 2r=2 n_{1,2}+1$. We are thus left with the single free parameter $\phi$, the squeezing angle. From Fig.\ref{fig:angolo}, we see that, for non-vanishing squeezing angles
{, and looking at the statistics of the $q_+$ EPR  quadrature,} a quantum advantage appears since the error probability curve can be lower than the corresponding classical bound. This is the main result of this work. In particular, we see that the advantage is maximized for a squeezing angle $\phi=\pi$ and can be shown to be monotonically increasing for $\phi\in (0,\pi]$. Fig.~\ref{fig:C_bound_n1} shows that no advantage is obtained when we set $\phi=0$ 
{, for measurements of the $q_+$ quadrature,} and it also shows the dependence of the quantum error probability on the mean number of photons in the noise input. As it could be expected, by increasing the mean number of photons in the input noise the quantum error probability increases. In the same way, the error probability increases when decreasing the number of repetitions of the experiment $N$ as it is clearly shown in Fig.~\ref{fig:C_bound_N}.

\begin{figure}
\centering
{\includegraphics[width=1.\columnwidth]{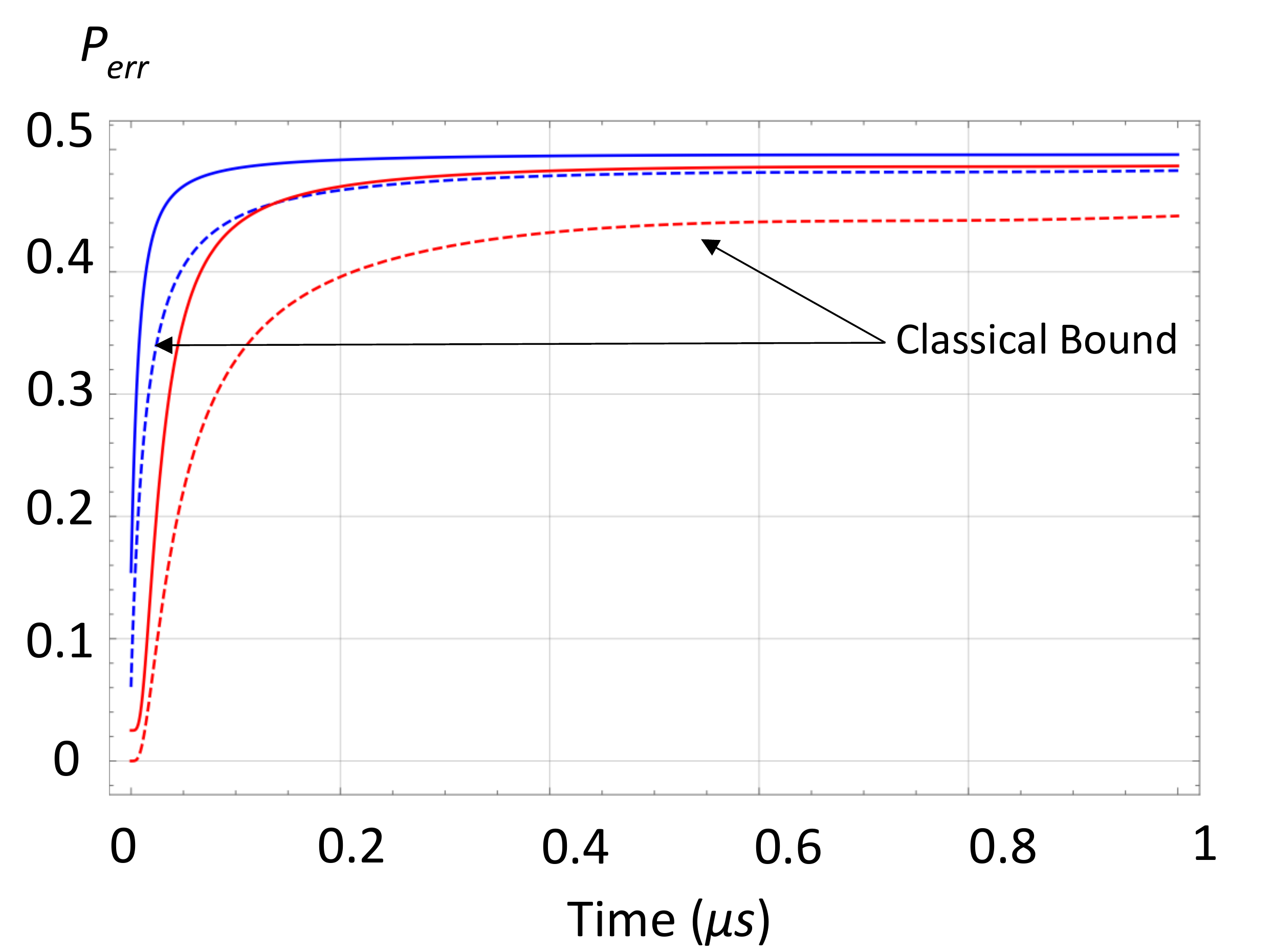}}
\caption{Error probabilities (solid lines) for a protocol in which we have input modes in a TMS state but classical measurements of the output modes. Here we consider local measurements of $x_{out_1}$ to perform the QHT. The dashed lines correspond to the classical bounds. Parameters values are $\phi=\pi$, $n_1=n_2=100$, and $\Delta=10^6$~Hz. The blue curves are obtained for $N=10$ and the red ones for $N=100$. This measurement scheme does not show any advantage in the form of a violation of the classical bound. }
\label{fig:TMS_C}
\end{figure}

\begin{figure}
\centering
{\includegraphics[width=0.5\textwidth]{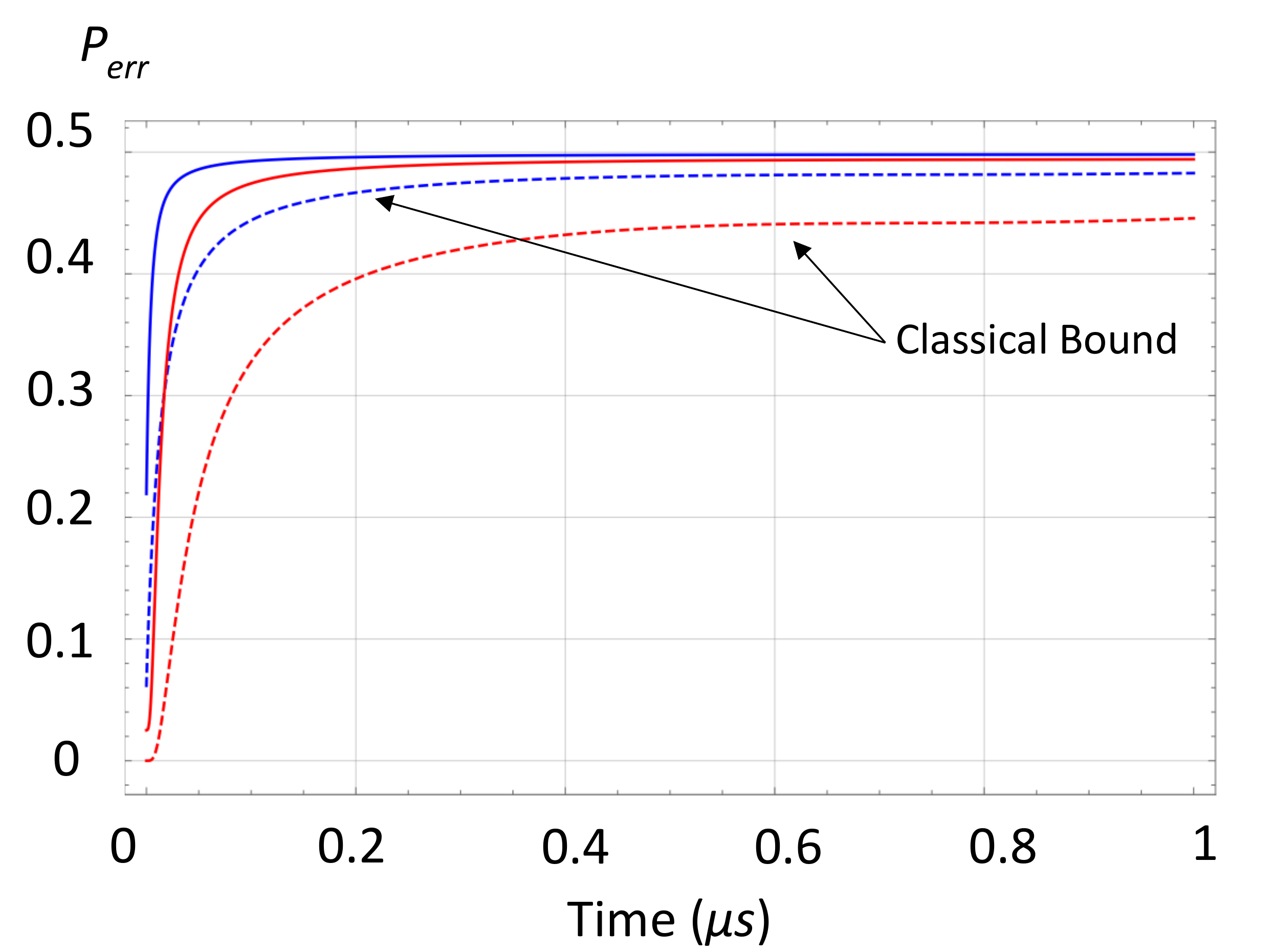}}
\caption{Error probabilities (solid lines) for a protocol in which we have the input modes in product of thermal states and EPR quadrature measurements for the output modes. Here we consider measurements of the quadrature $q_+$ to perform the QHT. The dashed lines correspond to the classical bounds. Parameters values are $\phi=\pi$, $n_1=n_2=100$, and $\Delta=10^6$~Hz. This measurement scheme does not show any advantage in the form of a violation of the classical bound.}
\label{fig:Thermal_EPR}
\end{figure}

\begin{figure}
\centering
{\includegraphics[width=1.\columnwidth]{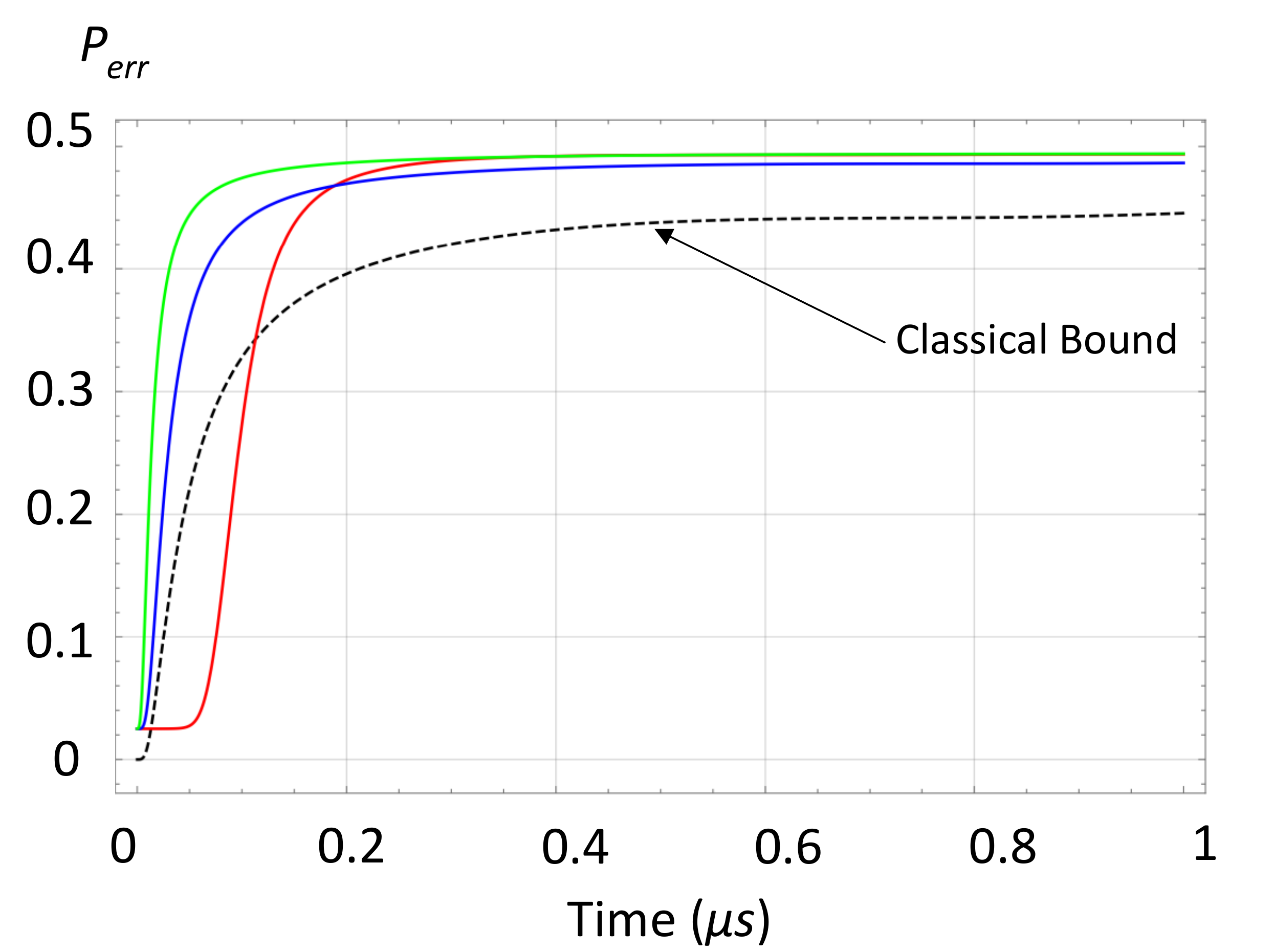}}
\caption{Comparison between different protocols: 1. Red curve: TMS light as input and EPR measurements of the output modes; 2. Blue curve: TMS light as input and classical (local) measurements of the output modes; 3. Green curve: thermal input noises and EPR measurements. Parameters are $\phi=\pi$, $n_1=n_2=100$, $N=100$, $\alpha=5\%$, and $\Delta=10^6$~Hz. The dashed line represents the corresponding classical bound. We consider the statistic of $x_{out_1}$ and $q_{+}$ from the local and EPR measurements, respectively. As already observed, the only protocol which is able to offer some advantage over the classical bound is the first, fully quantum one.}
\label{fig:confronto_schemi}
\end{figure}

For the sake of completeness, we examined also two additional cases: one is the combination of TMS light and local measurements and the other is the opposite case of classical thermal input noises and EPR measurements. Fig.~\ref{fig:TMS_C} and Fig.~\ref{fig:Thermal_EPR} show both these cases respectively. It is apparent that, when compared to their respective classical bounds, the error probabilities do not show any advantage even when fixing $\phi=\pi$ in the first case. This tells us that the quantum advantage as shown before depends on the combination of a quantum input and a quantum measurement strategy. Fig.~\ref{fig:confronto_schemi} shows the comparison between these last two protocols and the fully quantum one. 
{It should be noted that, in all the previous figures, neither the classical bounds nor the error probabilities vanish at the initial time.}

{When considering the error probabilities arising from the measurements of the EPR output quadratures, in all the reported figures we have shown the statistic of the EPR quadrature $q_+$. This is sufficient to demonstrate the quantum advantage by comparing error probability with the quantum bound. A similar performance would have been obtained by considering other output quadratures. For instance, if considering the statistics related to $q_-$, any quantum advantage is maximized for $\phi=0$.
This should not come at a surprise: the occurrence of an advantage when combining TMS input noise and EPR output measurements can be intuitively traced back to the fact that the two-modes squeezed input light allows quantum correlations of the output fields of the two cavity, which can be exploited in an EPR measurement. In line with the quantum reading protocol~\cite{Pirandola2011}, such correlations appear to be a quantum resource for HT inference. In this context, the dependence of the advantage on the squeezing angle can be qualitatively expected on the basis of the fact that -- depending on the parameters of the set-up -- the non-classical correlations between the output cavity modes that can enable the advantage can be accessed by measuring suitably rotated  phase-space quadratures.
\\
In the case of TMS input noise and local measurements, it is intuitive to understand that the quantum correlations established between the cavities' output modes cannot be exploited by a scheme based on local measurements. For instance, in Figs.~\ref{fig:TMS_C} and Fig.~\ref{fig:confronto_schemi} no advantage is shown. Analogously, in the case of Fig.~\ref{fig:Thermal_EPR}, where classical noise is teamed with EPR measurements, no advantage is expected. Indeed, as no quantum correlations between the output cavity modes can be present, the output mode of the second cavity is completely oblivious to the CSL mechanism affecting the mechanical mode in the first cavity. The mixing of the output cavity modes, entailed by the EPR measurement, can thus only additionally spoil the discrimination process as a noise source.\\
Finally, the quantum advantages we have found appear at short times and in a dynamical phase away from the steady-state. At long times we see from the previous figures that the advantage is not present anymore. This is analogous to what happens in certain quantum metrology schemes for open quantum systems~\cite{Gambetta} where, at long times, the effect of the dissipation is such that quantum properties are lost and so is the advantage.}

\begin{figure}
\centering
{\includegraphics[width=1.\columnwidth]{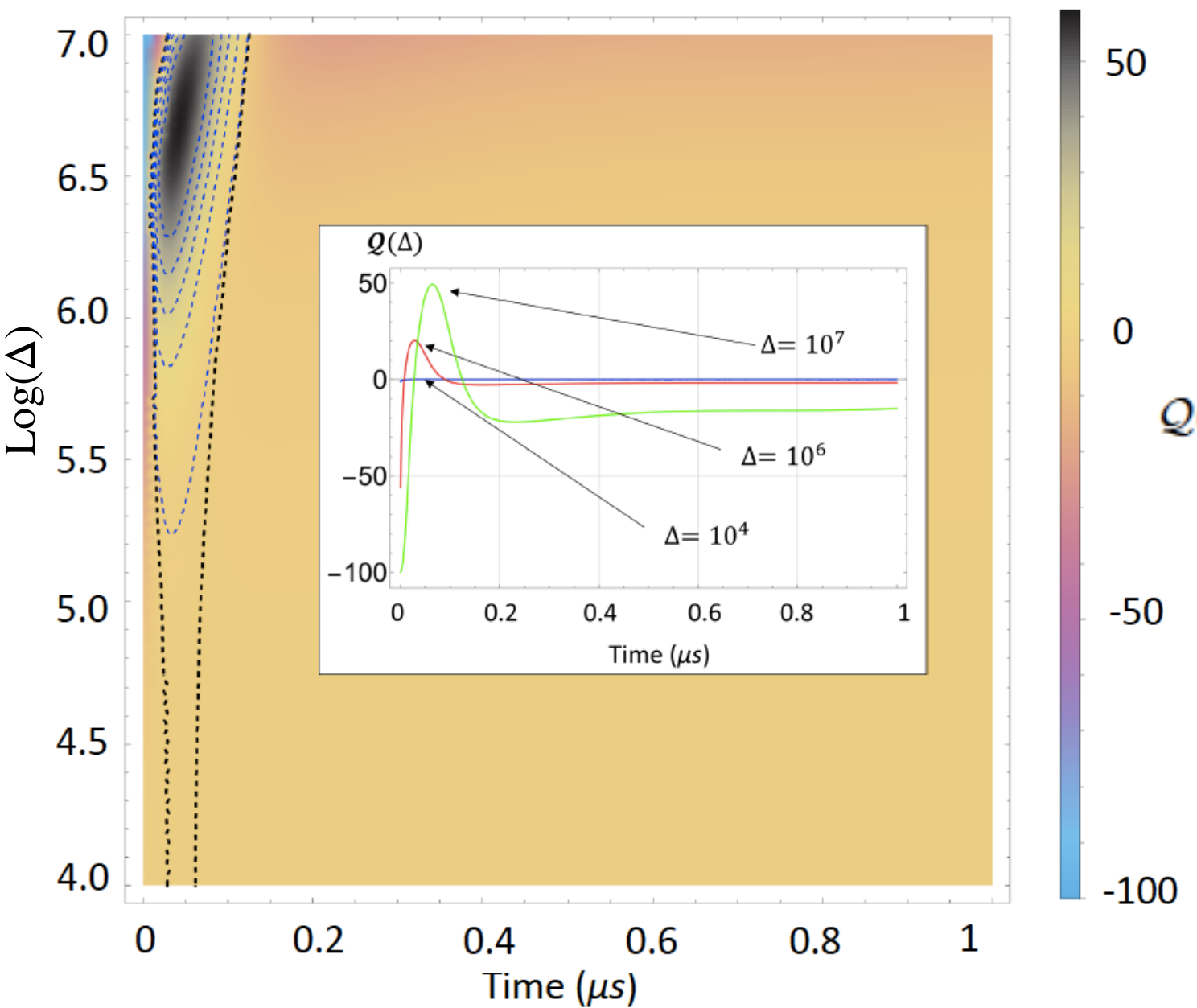}}
\caption{Quantum advantage for different values of the $\Delta$ parameter. The main figure shows a density plot of $\mathcal{Q}(\Delta,t)$ defined in Eq.~\eqref{violations}. Parameters are $n_1=n_2=100$, $N=10$, $\alpha=5\%$ and $\phi=\pi$, and we consider the statistic of measurements for the $q_{+}$ output quadrature. The black dashed contour separate the region of positive and negative $\mathcal{Q}$. The blue dashed contours are level lines in the region $\mathcal{Q}>0$, i.e., in the region in which a quantum advantage can be found. The inset shows three sample curves for $\Delta=\{10^4,10^6,10^7\}~$Hz in detail. It should be noted that, the case $\Delta=10^4$~Hz corresponds to the situation in which the CSL (or the unknown heating mechanism of the mechanical mode) rate is smaller than the thermal diffusion rate. We also observe that, for $\Delta=10^4$~Hz, despite the small quantum advantages, the quantum protocol considered delivers an error probability close to the classical bound at any time.}
\label{fig:deltavariabile}
\end{figure}

{To conclude this section, and} in view of the application of the QHT inference scheme presented here to collapse model, it is interesting to show that the quantum advantage persists if we vary the parameter $\Delta$. This can be seen in Fig.~\ref{fig:deltavariabile}, where it is shown the relative difference between the quantum error probability and the classical bound 
\begin{equation}\label{violations}
    \mathcal{Q}(\Delta)=100\frac{\mathcal{C}(n_1,n_2,t)-P_{err}(\phi,r,t)}{\mathcal{C}(n_1,n_2,t)+P_{err}(\phi,r,t)},
\end{equation}
for $\phi=\pi$, $n_1=n_2=100$, and $N=10$. This figure, and its inset, shows an advantage (regions of $\mathcal{Q}>0$) that is present at early times and extends on several order of magnitudes of $\Delta$. Remarkably, an advantage can still be found also when the unknown channel effect is sub-leading with respect to the thermal diffusion rate. An in-depth analysis of the possibilities offered by QHT for constraining CSL and related models is outside of the scope of the present work and will be part of a future investigation.

\section{Conclusions}\label{conclusions}
{The combination of QHT and optomechanical architectures opens the way to tests of fundamental physics and offers the possibility of quantum advantages over analogous classical strategies.

Following an approach inspired by the quantum reading framework, we have applied QHT to discriminate between two dynamical channels applied to an optomechanical system. Two hypotheses were formulated to describe the absence ($H_0$) or presence ($H_1$) of an additional dissipative mechanism, potentially due to the spontaneous localisation of the wavefunction of the mechanical resonator as predicted by the CSL model.

We compared two measurement strategies and we studied  the associated error probabilities to infer that there is an advantage when we use non-classical input noise states instead of classical resources. 

The classical scheme uses as input source two independent thermal states and it is combined with a direct measurement of the output modes. The quantum scheme employs a two-mode squeezed state and an EPR-like measurement. We have compared the error probabilities obtained from such schemes and the classical lower bound that can be obtained from the fidelity of the two-mode output state with classical thermal input noises.
While the error probabilities coming from the classical protocol are always greater than the classical bound, the same is not true for the quantum protocol error probability, which shows an advantage at finite times for some values of the squeezing angle.
Recently, in~\cite{2020arXiv200813580S} it was shown how squeezing entanglement offers an advantage for testing the CSL model in cold atom interferometric experiments. Moreover, we explored a large part of the parameter space of the CSL mechanism, showing that the advantage is widespread.

In the framework of collapse models, this study offers a starting point to future analysis aimed to restrict the range of still untested parameters characterizing CMs. More in general, we have proposed a versatile scheme that could be implemented and applied to different systems in view of exploring other fundamental physics mechanisms.

}

\acknowledgments
MMM and MP acknowledge support from the EU H2020 FET Project TEQ (Grant No. 766900). AB acknowledges the MSCA project pERFEcTO (Grant No. 795782). 
SP acknowledges funding from the European Union’s Horizon 2020 Research and Innovation Action under grant agreement No. 862644 (FET-OPEN project: Quantum readout techniques
and technologies, QUARTET).
MP is supported by the DfE-SFI Investigator Programme (grant 15/IA/2864), the Royal Society Wolfson Research Fellowship
(RSWF\textbackslash R3\textbackslash183013),
the Royal Society International Exchanges Programme
(IEC\textbackslash R2\textbackslash192220),
the Leverhulme Trust Research Project Grant (grant nr.~RGP-2018-266), and the UK EPSRC (grant nr.~EP/T028106/1). This research was partially supported by COST Action CA15220 ``Quantum Technologies in Space''.

\bibliographystyle{apsrev4-1.bst}
\bibliography{references} 

\end{document}